\documentclass{optica-article}

\journal{opticajournal} 

\usepackage{orcidlink}
\articletype{Research Article}
\usepackage{svg}
\usepackage{booktabs}
\usepackage{makecell}

\begin{document}

\title{Dual-Pilot-Assisted Frequency and Phase Recovery in Experimental CV-QKD}

\author{Christiano M. S. Nascimento \authormark{1,2*} \orcidlink{0009-0000-2160-8742}, Artur A. Matoso \authormark{1} \orcidlink{0000-0003-4039-0544}, Gustavo C. Amaral \authormark{2,3} \orcidlink{0000-0001-6071-7458} and Guilherme P. Temporão\authormark{2} \orcidlink{0000-0002-5244-3394}}

\address{\authormark{1}QuIIN – Quantum Industrial Innovation, EMBRAPII CIMATEC Competence Center in Quantum Technologies, SENAI CIMATEC, Av. Orlando Gomes 1845, Salvador, 41650-010, BA, Brazil\\
\authormark{2}NITeQ, Department of Electrical Engineering, Pontifical Catholic University of Rio de Janeiro, 22451-900 Rio de Janeiro, RJ, Brazil\\
\authormark{3}Quantum Technology Department, TNO—The Netherlands Organization for Applied Scientific Research, 2628CK Delft, The Netherlands}

\email{\authormark{*}christiano.moreira@fieb.org.br} 


\begin{abstract*} 
Continuous-variable quantum key distribution (CV-QKD) offers high secret-key rates and compatibility with existing optical communication infrastructure. Architectures that use a locally generated local oscillator (LLO) eliminate the need to transmit a strong optical reference through the quantum channel, reducing associated security vulnerabilities. However, independent transmitter and receiver lasers introduce frequency offsets and phase fluctuations, making accurate carrier-phase recovery one of the main challenges in LLO-based CV-QKD systems. Conventional implementations commonly rely on a single pilot tone for phase recovery. Here, we demonstrate a Gaussian-modulated CV-QKD system that uses two frequency-multiplexed pilot tones for carrier-frequency-offset estimation and phase-noise compensation over a 12 km optical fiber spool. We present a complete digital signal processing (DSP) chain integrating CAZAC-based frame synchronization, global and local frequency-offset estimation, and dual-pilot-assisted carrier-phase recovery. The experimental system employs RF-heterodyne detection with fully independent laser sources and processes logical frames containing approximately $10^6$ coherent states. Using experimental channel parameters, we evaluate security under a trusted-receiver model, yielding average modeled secret-key rates of $14.98$~Mbit/s and $8.17$~Mbit/s in the asymptotic and finite-size parameter-estimation regimes, respectively.
Compared with single-pilot processing, the proposed dual-pilot DSP chain reduced the average input-referred excess noise by 28.1\% and 17.4\% relative to pilots 1 and 2, respectively, demonstrating the benefits of jointly exploiting both references for frequency and phase recovery in experimental LLO CV-QKD.
These results demonstrate the feasibility of an integrated dual-pilot-assisted DSP architecture for LLO-based Gaussian CV-QKD and provide a framework for practical metropolitan quantum communication systems.
\end{abstract*}

\section{Introduction}
\label{sec1}

Continuous-variable quantum key distribution (CV-QKD) enables two legitimate parties, Alice and Bob, to establish secret keys by encoding information in the quadratures of optical fields \cite{hajomer2024long,grosshans2001continuous,weedbrook2004quantum}. In Gaussian-modulated protocols, Alice prepares coherent states whose amplitude and phase quadratures follow Gaussian distributions, while Bob performs homodyne or heterodyne detection \cite{laudenbach2018continuous,jouguet2013experimental,pietri2024qosst}. By publicly disclosing a subset of their correlated measurements, Alice and Bob estimate the quantum-channel parameters and bound the information potentially accessible to an eavesdropper (Eve). The remaining data undergo classical post-processing, including information reconciliation and privacy amplification, to extract a secret key \cite{laudenbach2018continuous,weedbrook2012gaussian}.

Early CV-QKD implementations transmitted a strong local oscillator (LO) alongside the quantum signal, providing a common phase reference that mitigates relative laser phase fluctuations \cite{juvencio2026towards,grosshans2001continuous,weedbrook2012gaussian,wang2024high,weerasinghe2023practical}. However, transmitting the LO introduces potential security vulnerabilities, since an adversary may manipulate the reference field and compromise parameter estimation \cite{do2025side}. Locally generated local oscillator (LLO) architectures address these vulnerabilities by employing an independent laser at Bob and transmitting optical references for phase recovery \cite{hajomer2024long,hajomer2025coexistence,laudenbach2018continuous}. Nevertheless, independent laser sources introduce carrier-frequency offsets (CFO) and phase fluctuations that must be compensated through digital signal processing (DSP). Imperfect frequency and phase recovery can increase excess noise, thereby reducing the achievable secret-key rate.

Several approaches have addressed these challenges, including time-multiplexed reference symbols and frequency-multiplexed pilot tones \cite{hajomer2024long,laudenbach2018continuous}. The open-source QOSST framework employs two frequency-multiplexed pilots for sampling-clock synchronization while primarily relying on one pilot for carrier-frequency and phase recovery \cite{pietri2024qosst}. More recently, Liao \textit{et al.} investigated multi-pilot phase recovery using phase averaging and Wiener filtering, demonstrating the potential benefits of combining multiple phase references \cite{liao2025simple}. Building on these developments, we experimentally demonstrate a Gaussian-modulated LLO CV-QKD system integrating CAZAC-based frame synchronization, global and subframe-wise CFO estimation, and coherent dual-pilot phase recovery. The system operates at 156.25~MBaud over 12~km of optical fiber using RF-heterodyne detection with independent lasers. We evaluate 120 experimental frames, each containing approximately $10^6$ coherent states, and directly compare single-pilot and coherent dual-pilot processing using identical acquisitions. The proposed dual-pilot DSP chain reduces the average input-referred excess noise by 28.1\% and 17.4\% relative to pilots 1 and 2, respectively, while enabling positive modeled secret-key rates under a trusted-receiver security model.

\section{Methodology}
\label{sec:methodology}

At the receiver, the pilot frequencies are identified from the acquired signal spectrum. Because of the frequency difference between Alice's and Bob's lasers, the received pilot frequencies are shifted relative to their transmitted values. An individual CFO estimate is obtained from each pilot as

\begin{equation}
    \Delta f_k = \hat{f}_{p,k}^{\mathrm{Rx}} - f_{p,k}^{\mathrm{Tx}}\text{,}
    \qquad k \in \{1,2\}\text{,}
\end{equation}

\noindent where $\hat{f}_{p,k}^{\mathrm{Rx}}$ and $f_{p,k}^{\mathrm{Tx}}$ denote the measured and transmitted pilot frequencies, respectively. The global CFO is then estimated by averaging both contributions:

\begin{equation}
    \widehat{\Delta f}_{\mathrm{global}} = \frac{\Delta f_1+\Delta f_2}{2}\text{.}
    \label{eq:global_cfo}
\end{equation}

We compensate for the estimated frequency offset by applying a complex rotation to the acquired signal. To account for frequency variations throughout the acquisition, the corrected waveform is subsequently divided into subframes. Within each subframe, both pilot frequencies are re-estimated, and the residual CFO is calculated using the same dual-pilot procedure. This two-stage approach compensates for the dominant frequency offset while allowing local tracking of frequency variations during transmission.

Following frequency compensation, the two pilot tones are independently translated to complex baseband and isolated using low-pass finite impulse response (FIR) filters. The recovered pilot signals are expressed as

\begin{equation}
    p_k(t)=
    \operatorname{LPF}
    \left\{
    r(t)e^{-j2\pi f_{p,k}t}
    \right\}\text{,}
    \qquad k\in\{1,2\}\text{,}
    \label{eq:pilot_recovery}
\end{equation}

where $r(t)$ represents the frequency-corrected received signal and $f_{p,k}$ denotes the nominal pilot frequency after CFO compensation. Before combining the recovered references, we remove the known initial phases of the transmitted pilots, $\phi_{k,0}$. The resulting complex signals are then coherently combined according to

\begin{equation}
    p_{\mathrm{c}}(t)= p_1(t)e^{-j\phi_{1,0}} + p_2(t)e^{-j\phi_{2,0}}\text{.}
    \label{eq:dual_pilot_combination}
\end{equation}

Unlike conventional phase-domain averaging, where each pilot's phase is estimated independently before combination, Eq.~\eqref{eq:dual_pilot_combination} combines them directly in the complex domain. Consequently, each pilot's contribution depends on its recovered amplitude and relative phase. When both references exhibit similar amplitudes and closely matched phases, the procedure approaches conventional phase averaging. However, with unequal pilot amplitudes or different phase perturbations, the resulting phase estimate generally differs from the arithmetic mean of the individual pilot phases.

The instantaneous phase of the combined reference is subsequently extracted, unwrapped, and temporally smoothed:

\begin{equation}
    \widehat{\phi}(t)= \mathcal{S} \left\{ \text{unwrap}
    \left[ \arg\left(p_{\mathrm{c}}(t)\right) \right] \right\}\text{,}
    \label{eq:dual_pilot_phase}
\end{equation}

\noindent where $\mathcal{S}\{\cdot\}$ denotes a temporal smoothing operator employed to suppress high-frequency fluctuations in the phase estimate; assuming both pilots experience predominantly common laser-induced phase noise, their coherent combination provides a single phase reference to compensate the received quantum signal. The estimated phase is removed through

\begin{equation}
    r_{\mathrm{corr}}(t)=
    r(t)e^{-j\widehat{\phi}(t)}\text{.}
    \label{eq:phase_compensation}
\end{equation}

The proposed processing strategy therefore combines global and subframe-wise CFO estimation with coherent dual-pilot phase recovery. Although related multi-pilot phase-combination techniques have been investigated in optical communications, the present implementation exploits the same two pilot tones across the frequency- and phase-recovery stages of an experimental LLO-CV-QKD receiver \cite{hajomer2024long,liao2025simple,pietri2024qosst}. The resulting phase-corrected signal is subsequently processed through matched filtering and symbol extraction for quantum-channel characterization and secret-key-rate estimation.  For comparison, the same experimental acquisitions were also processed using each pilot individually for CFO estimation and phase recovery, while retaining both transmitted pilot tones and the remaining receiver configuration.

\section{LLO CV-QKD Experimental Setup}
\label{sec:II}

\begin{figure}[htpb]
    \centering
    \includegraphics[width=\textwidth]{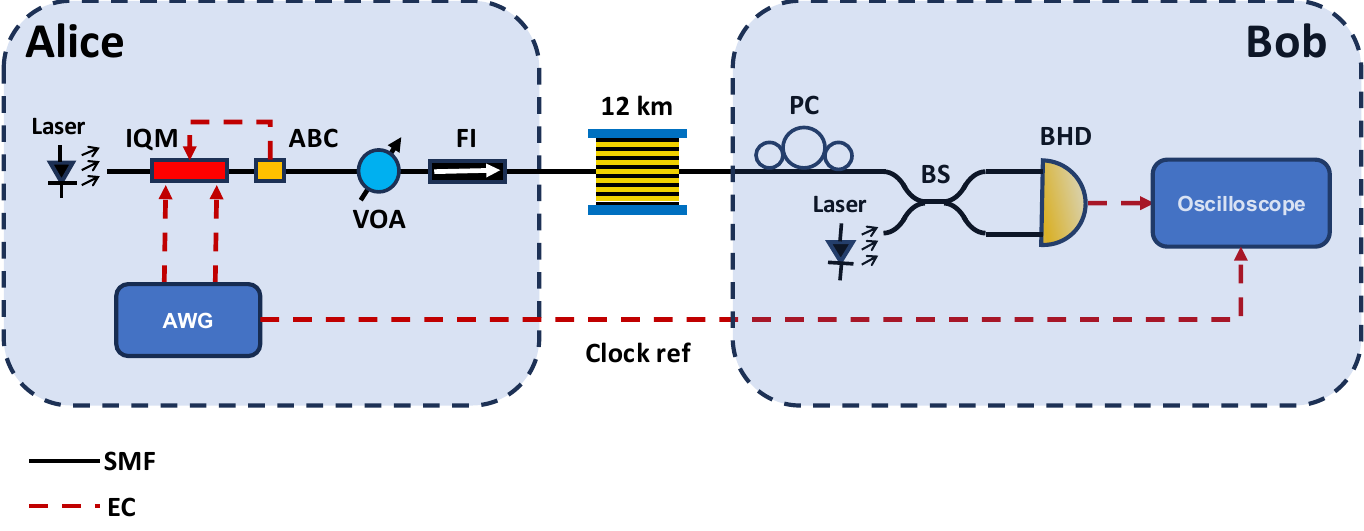}
    \caption{Experimental setup of LLO CV-QKD. Alice’s station consists of a continuous-wave (CW) laser operating at 1550 nm and an in-phase and quadrature modulator (IQM) with an automatic bias controller (ABC) driven by an arbitrary waveform generator (AWG) with 16-bit resolution and a sampling rate of 3.125 GSa/s. A variable optical attenuator (VOA) is placed after the IQ modulator (at the exit of Alice's station) to adjust the optical power and enforce the few-photon regime. A 12 km single-mode fiber (SMF) constitutes the quantum channel. At Bob’s station, a polarization controller (PC) allows one to align the polarization of the incoming signal to the LO signal generated from an independent CW laser; both overlap at a balanced beam splitter. The differential signal was detected and digitized using a balanced detector (BD), followed by a 12-bit oscilloscope with a matched sampling rate of 3.125 GSa/s. EC: electrical connection.}
    \label{fig:setup}
\end{figure}

Fig. \ref{fig:setup} shows the experimental setup of the LLO CV-QKD laboratory link including the optical layout and the electro-optical devices necessary to perform the Gaussian-modulated coherent-state protocol.

At the sender, Alice, a continuous-wave (CW) laser with a narrow linewidth of 1 kHz operating at 1550 nm was used as the optical carrier. The coherent states were prepared by driving an IQ modulator with the output from a 16-bit arbitrary waveform generator (AWG) with two channels operating at a sampling rate of 3.125 GSa/s. The IQ modulator was operated in single-sideband mode by digitally frequency-shifting the transmitted signal and controlling the direct-current bias voltages with an automatic bias controller (ABC). A variable optical attenuator (VOA) was introduced after the IQ modulator to adjust the modulation variance of the thermal state. Furthermore, a Faraday isolator at the exit of Alice's station; this helps mitigate back-reflections from the optical fiber channel and is a countermeasure against so-called Trojan-horse attacks \cite{do2025side}. In this implementation, the IQ encoding was chosen based on a pseudo-random Gaussian sequence generated in a standard personal computer (PC) and serves as a proof of concept for the operation of the experiment rather than actual secret key material generation \cite{grosshans2001continuous, laudenbach2018continuous}; in the future, this functionality can be seamlessly replaced by a quantum random number generator (QRNG) without altering the experimental setup.

The signal was sent through a quantum channel consisting of a 12 km single-mode fiber, with a measured average attenuation of 0.185 dB/km at 1550 nm, i.e., a total channel loss of 2.22 dB. At the receiver, Bob implements RF heterodyne detection to evaluate both quadratures simultaneously \cite{hajomer2024long, pietri2024qosst}. This is accomplished by overlapping Bob's CW laser -- also operating at a center wavelength of 1550 nm with a linewidth of 1 kHz (the LLO) -- with the incoming signal from Alice. The frequency difference between Alice’s and Bob’s lasers was approximately 800 MHz. The two beams at the balanced beam splitter inputs are aligned in polarization with manual polarization controllers and, after combination into a single spatial mode, impinge on a balanced detector with a bandwidth of 1 GHz that detects the interference pattern. The latter was digitized using a 12-bit oscilloscope with a sampling rate of 3.125 GSa/s, matched to Alice's sampling rate and with its internal clock synchronized to the AWG's clock, and recorded for offline DSP and post-processing.

\section{Results}

To evaluate the system performance, we adopted a trusted-device security model, in which part of the system loss and noise is assumed to be inaccessible to Eve \cite{hajomer2024long, scarani2009security}. Tab. \ref{tab:experimental_parameters} summarizes the average experimental parameters obtained from 120 frames and used in the secret-key calculations in the asymptotic regime.

\begin{table}[htbp]
    \centering
    \caption{Average experimental parameters and results under the trusted-device security model over 120 frames.}
    \label{tab:experimental_parameters}

    \setlength{\tabcolsep}{5pt}
    \renewcommand{\arraystretch}{1.2}

    \begin{tabular}{cccccccc}
        \toprule
        \makecell{$R_s$ (Mbaud)}
        &
        \makecell{$V_{mod}$ (SNU)}
        &
        \makecell{$\eta_{\mathrm{Bob}}$}
        &
        \makecell{$v_{\mathrm{el}}$ (mSNU)}
        &
        \makecell{$T$}
        &
        \makecell{$\xi_{B}$ (mSNU)}
        &
        \makecell{FER}
        &
        \makecell{$\beta$ (\%)}
        \\
        \midrule

        156.25 & 0.96 & 0.80 & 44.29 & 0.5855 & 3.63 & 0.276 & 93.54 \\

        \bottomrule
    \end{tabular}
\end{table}

For each analyzed logical frame, Alice prepared approximately $10^6$ coherent states. The average modulation variance $V_{mod}$ was $0.96$ SNU. The quantum channel exhibited an average untrusted transmittance of $T = 0.5855$ and excess noise of $\xi_{B}=3.63$ mSNU at Bob in the asymptotic regime. This excess noise can be attributed to different sources, e.g., phase noise from the laser linewidth \cite{silva2026limits, laudenbach2018continuous}. The detector's average electronic noise was $44.29$ mSNU, while detection efficiency was $0.80$. The clearance of the vacuum noise over the electronic noise was about 13 dB. To calibrate $V_{mod}$ of the thermal state, we performed back-to-back measurements, in which Alice and Bob were connected through a short fiber patch cord.

At Bob's station, the acquired signals are processed using the dual-pilot-assisted DSP methodology described in Sec.~\ref{sec:methodology}, combined with CAZAC-based frame synchronization and RRC-matched filtering. Vacuum and electronic-noise measurements undergo the same frequency correction, filtering, and sampling procedures to ensure consistent normalization and parameter estimation.

After the DSP stage, the recovered quantum symbols undergo classical post-processing, including information reconciliation, parameter estimation, and privacy amplification. Information reconciliation is based on multidimensional reconciliation combined with multi-edge-type low-density parity-check (MET-LDPC) error-correcting codes \cite{mani2021multiedge} with a rate of 0.131. This procedure enables Alice and Bob to correct discrepancies between their correlated measurement sequences through reverse reconciliation. However, the reconciliation performance depends on the experimental signal-to-noise ratio (SNR), which may vary during transmission because of phase-recovery imperfections, polarization fluctuations, and other experimental instabilities \cite{hajomer2024long,nascimento2025passive}.

We estimate channel parameters using the recovered quantum symbols to determine the channel transmittance and excess noise, which we then use to bound Eve's accessible information via the Holevo quantity. We evaluate each acquisition frame independently, without concatenating multiple frames into a larger security block. We consider both asymptotic and finite-size parameter-estimation regimes. In the asymptotic regime, we use the estimated channel parameters directly in the security analysis. In the finite-size regime, we construct conservative confidence intervals to account for statistical uncertainties arising from the limited number of available symbols. Specifically, we determine the lower confidence bound on the channel transmittance using the Student's t-distribution. At the same time, we obtain the upper confidence bound on the excess noise using the chi-squared distribution \cite{juvencio2026towards}. These worst-case parameters are then used to evaluate the Holevo bound and estimate the achievable secret-key rate under a trusted-receiver model. Privacy amplification, when performed on successfully reconciled and verified bit sequences, employs universal hashing based on Toeplitz matrices to reduce the information potentially available to Eve \cite{laudenbach2018continuous,juvencio2026towards}. All DSP and classical post-processing operations are carried out offline.

The secret key rate (SKR) is given by:

\begin{equation}
    \mathrm{SKR} = R_s (1-\mathrm{FER}) [\beta I_{\text{AB}} - \chi_{\text{BE}}],
    \label{eq:skr_finite}
\end{equation}

\noindent where $I_{\text{AB}}$ is the mutual information between Alice and Bob, and $\chi_{\text{BE}}$ is the Holevo bound. Finite-size parameter estimation employs a failure probability of $\epsilon_{\mathrm{PE}}=10^{-12}$. We emphasize that these security parameters do not, by themselves, constitute a universally composable security analysis.

\begin{figure}[htpb]
    \centering
    \includegraphics[width=\textwidth]{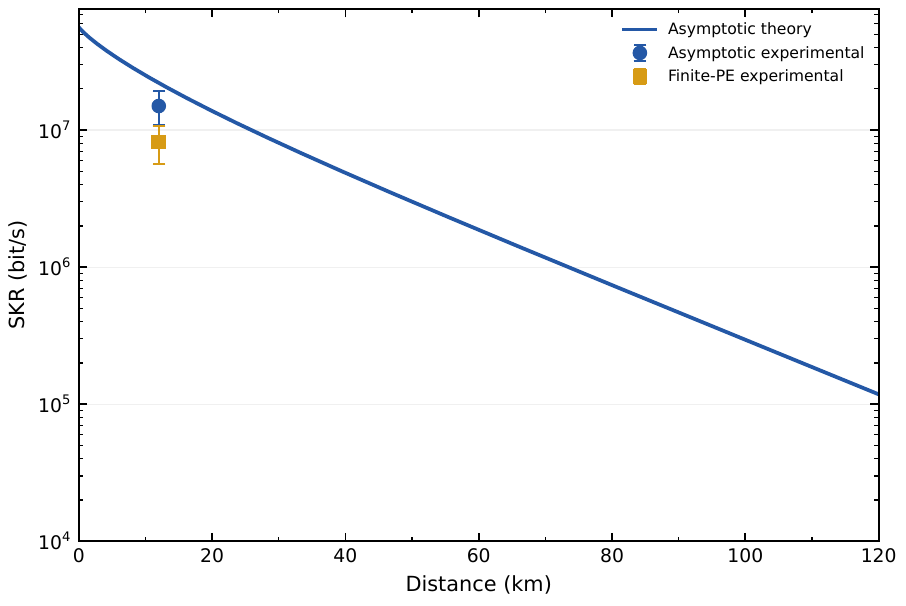}
    \caption{Asymptotic SKR versus fiber distance for an attenuation of 0.185~dB/km, normalized to the maximum experimental asymptotic SKR at 12~km. Experimental markers show the mean asymptotic and finite-PE SKRs, with error bars representing standard deviations across 120 frames.}
    \label{fig:SKR}
\end{figure}

Fig. \ref{fig:SKR} presents the numerically simulated asymptotic SKR as a function of transmission distance, together with the experimental results obtained over 12 km of optical fiber. Considering 120 experimental frames, the system achieved an average asymptotic SKR of $14.98 \pm 4.18$ Mbit/s and a finite-size parameter estimation (finite-PE) SKR of $8.17 \pm 2.53$ Mbit/s, where the uncertainties represent the standard deviation across the experimental frames. The maximum experimental asymptotic SKR reached 22.08 Mbit/s, which was used to normalize the theoretical curve at the experimental transmission distance. The numerical model incorporates the measured system parameters, including detector efficiency, electronic noise, excess noise, reconciliation efficiency, and frame error rate, while assuming these parameters remain constant as the channel attenuation increases. Under these assumptions, the model provides an estimate of the achievable asymptotic SKR over transmission distances of up to 120 km. Although this extrapolation does not account for distance-dependent experimental impairments, the results indicate the potential of the proposed LLO CV-QKD architecture for metropolitan fiber networks, motivating its future integration into the \textit{Rede Rio Quântica Network} in Rio de Janeiro, Brazil.
 
\section{Discussion}
\label{sec:III}

\begin{figure*}[t]
    \centering
    \includegraphics[width=\textwidth]{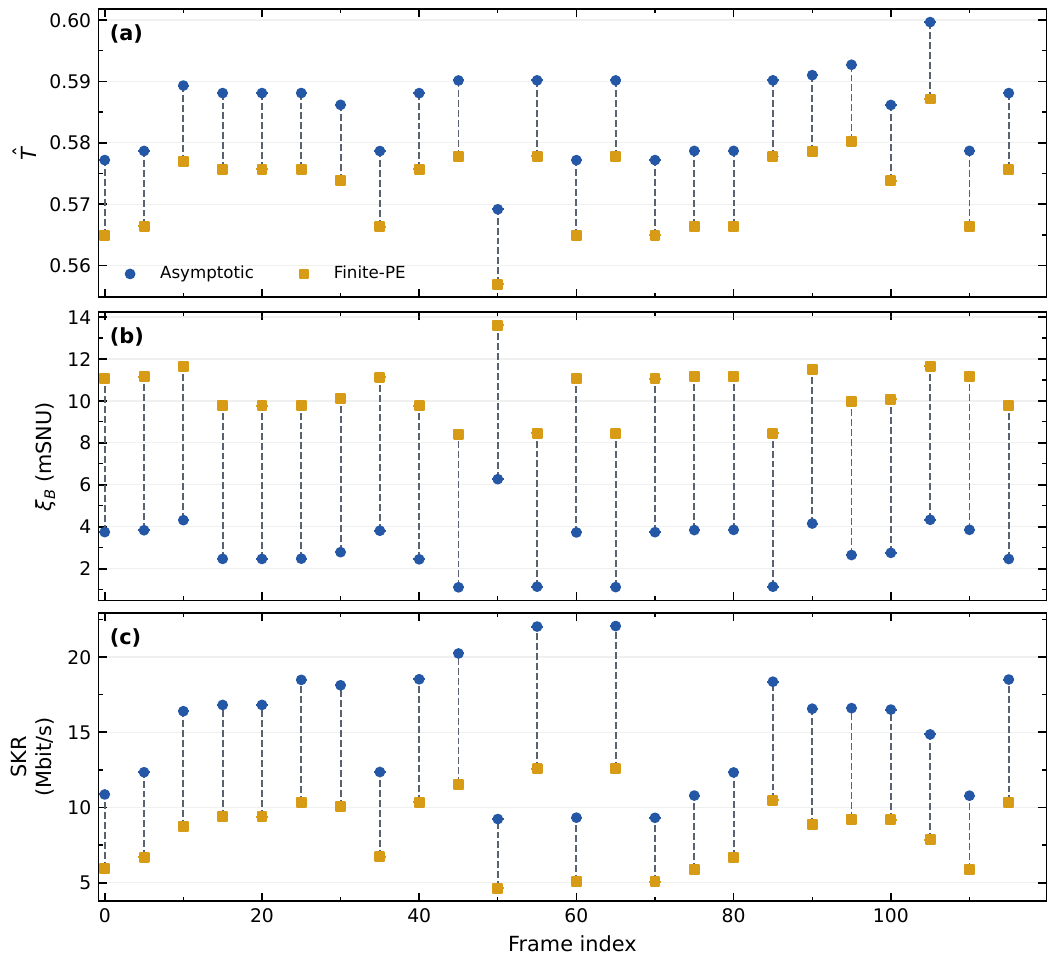}
    \caption{Frame-resolved experimental performance over the 12-km fiber link. (a) Estimated channel transmittance $\hat{T}$ and its conservative finite-size parameter-estimation lower bound $T_{\min}$. (b) Estimated excess noise referred to Bob, $\xi_B$, and its finite-PE upper bound $\xi_{B,\max}$. (c) Modeled secret-key rates obtained from asymptotic and finite-PE parameter estimates. For clarity, we show every fifth frame. The dashed segments connect point estimates to conservative finite-PE bounds and do not represent standard deviations. Finite-PE results do not constitute a complete composable finite-key security proof.}
    \label{fig:frame_stability}
\end{figure*}

Fig.~\ref{fig:frame_stability} further investigates the stability of the experimental system by presenting the estimated channel transmittance, excess noise, and secret-key rates over the acquired frames. We compare the asymptotic estimates with conservative bounds from finite-size parameter estimation, allowing us to evaluate the statistical uncertainty associated with finite experimental blocks. As shown in Fig.~\ref{fig:frame_stability}(a), the estimated transmittance remains relatively stable throughout the measurements, with most values concentrated around 0.58. Similarly, Fig.~\ref{fig:frame_stability}(b) shows that the excess noise referred to Bob remains at the millishot-noise-unit level, with several frames exhibiting asymptotic estimates below 3~mSNU. These observations indicate that the receiver maintains relatively stable channel estimation and low excess-noise levels throughout the acquisition sequence.

Fig.~\ref{fig:frame_stability}(c) shows the corresponding secret-key-rate results. Considering 120 experimental frames, the system achieved an average asymptotic SKR of $14.98 \pm 4.18$~Mbit/s and a finite-PE SKR of $8.17 \pm 2.53$~Mbit/s, where the uncertainties represent the standard deviations across the experimental frames. The maximum experimental asymptotic SKR reached 22.08~Mbit/s. Although finite-size parameter estimation noticeably reduces the estimated secret-key rate, we obtained positive finite-PE SKR values across all displayed measurements. The difference between asymptotic and finite-PE estimates reflects the statistical penalty of finite experimental blocks, particularly through the conservative lower bound on channel transmittance and the upper bound on excess noise. These results show that the implemented receiver maintains positive modeled secret-key rates under the considered experimental conditions.

To quantify the effect of dual-pilot processing, we reprocessed the same 120 acquisitions using either pilot individually or their coherent combination. As shown in Table~\ref{tab:pilot_processing_comparison}, the dual-pilot configuration reduced the mean input-referred excess noise by 28.1\% and 17.4\% relative to pilots 1 and 2, respectively, and yielded the lowest excess noise in all 120 paired frames. Because the CFO and phase-recovery procedures differ between configurations, this improvement reflects the full dual-pilot DSP chain rather than the coherent phase combination alone.

\begin{table}[htbp]
    \centering
    \caption{Experimental excess-noise comparison between single-pilot and coherent dual-pilot processing over 120 frames.}
    \label{tab:pilot_processing_comparison}
    \renewcommand{\arraystretch}{1.25}
    \setlength{\tabcolsep}{7pt}
    \begin{tabular}{lcccc}
        \hline
        & \multicolumn{2}{c}{$\xi_{\mathrm{in}}$ (mSNU)}
        & \multicolumn{2}{c}{$\xi_B$ (mSNU)} \\
        \cline{2-3} \cline{4-5}
        Method & Mean & Std. dev. & Mean & Std. dev. \\
        \hline
        Pilot 1 & 21.65 & 7.80 & 5.03 & 1.77 \\
        Pilot 2 & 18.85 & 7.57 & 4.39 & 1.72 \\
        \hline
        \textbf{This work} & \textbf{15.56} & \textbf{7.61} & \textbf{3.63} & \textbf{1.74} \\
        \hline
    \end{tabular}
\end{table}

To contextualize the proposed system's experimental performance, Table~\ref{tab:cvqkd_literature_comparison} compares it with representative CV-QKD demonstrations reported in the literature. Jain \textit{et al}.~\cite{jain2022practical} demonstrated practical CV-QKD with composable security. In contrast, Piétri \textit{et al.}~\cite{pietri2024qosst} introduced the QOSST platform, employing two frequency-multiplexed pilot tones for synchronization and carrier recovery, with one pilot used for phase tracking. More recently, Liao \textit{et al.}~\cite{liao2025simple} investigated multiple-frequency pilots for phase recovery in CV-QKD, showing that combining phase estimates from different pilots and applying Wiener filtering can improve phase-recovery performance. Other experimental demonstrations have explored multicarrier transmission and neural-network-assisted phase compensation to improve the achievable transmission distance and secret-key rate.

\begin{table*}[t]
    \centering
    \caption{Comparison of experimentally reported excess noise in representative CV-QKD demonstrations. The excess noise is expressed in mSNU and referred either to the channel input ($\xi_{\mathrm{in}}$) or to Bob's side ($\xi_B$), according to the convention adopted in each reference. Different modulation formats, detection models, and noise normalizations prevent a direct performance ranking.}
    \label{tab:cvqkd_literature_comparison}
    \renewcommand{\arraystretch}{1.30}
    \setlength{\tabcolsep}{6pt}
    \normalsize
    \begin{tabular*}{\textwidth}{@{\extracolsep{\fill}}lccccc@{}}
        \hline
        Reference & Pilots & $R_s$ (MBaud) & $L$ (km) & $\xi_{\mathrm{in}}$ (mSNU) & $\xi_B$ (mSNU) \\
        \hline
        \cite{pan2022experimental} & 1 & 1000 & 50.59 & 20.0 & -- \\
        \cite{jain2022practical} & 1 & 100 & 20 & -- & 10.4$^{\mathrm{a}}$ \\
        \cite{ruiz2024low} & 1 & 100 & 25 & 91.0 & -- \\
        \cite{pietri2024qosst} & 2 & 100 & 10$^{\text{b}}$ & -- & 7.6 \\
        \cite{pietri2024qosst} & 2 & 100 & 25 & -- & 7.2 \\
        \cite{pietri2024experimental} & 2 & 100 & 23 & -- & 9.0 \\
        \cite{liao2025simple} & 2 & 1000 & 20 & -- & 20.04 \\
        \cite{pan2025long} & 1 & 1000 & 126.56 & 19.0 & -- \\
        \hline
        \textbf{This work} & \textbf{2} & \textbf{156.25} & \textbf{12} & $\boldsymbol{15.56 \pm 7.61}$ & $\boldsymbol{3.63 \pm 1.74}$ \\
        \hline
    \end{tabular*}
    \vspace{4pt}
    \parbox{\textwidth}{\footnotesize $^{\mathrm{a}}$ Converted from 5.2~mPNU of untrusted noise at the channel output using the normalization reported in the original work; this is not necessarily equivalent to detector-referred excess noise. $^{\mathrm{b}}$ Equivalent attenuation obtained using a variable optical attenuator. $^{\mathrm{c}}$ Worst-case excess-noise estimate rather than an asymptotic mean. Dashes indicate values not established in the corresponding convention. For this work, uncertainties are sample standard deviations across 120 experimental frames.}
\end{table*}

Among the reported demonstrations, Liao \textit{et al}.~\cite{liao2025simple} is particularly relevant because it explicitly investigates phase recovery using multiple frequency-multiplexed pilots in CV-QKD. In their approach, they independently estimate each pilot's phase before combining the estimates through arithmetic averaging. In contrast, the receiver presented here coherently combines the individually downconverted and phase-aligned complex pilot fields before extracting the common phase estimate. Although these approaches are mathematically equivalent under ideal conditions with equal pilot amplitudes and sufficiently aligned phases, differences may arise with unequal pilot amplitudes, independent noise contributions, or relative phase fluctuations. Complex pilot combining has been explored in classical coherent optical communication systems; therefore, the technique itself is not claimed as a fundamentally new phase-estimation principle.

This work's main contribution is the experimental integration of dual-pilot-assisted frequency and phase recovery with matched filtering, reconciliation, and parameter estimation in a CV-QKD receiver. Operating at 156.25~MBaud, the system achieved an average input-referred excess noise of $15.56\pm7.61$~mSNU and positive modeled secret-key rates under asymptotic and finite-PE estimates. For comparison, Pan \textit{et al.}~\cite{pan2025long} reported 19~mSNU at 1~GBaud over 126.56~km, although different experimental conditions prevent a direct performance comparison. Because laser phase diffusion per symbol depends on the ratio of linewidth to symbol rate \cite{hajomer2024long,de2025digital}, future work will explore dual-pilot recovery in the GBaud regime and compare coherent pilot combining with conventional phase averaging. Finally, the finite-PE analysis incorporates conservative channel-parameter bounds but does not constitute a complete composable security proof.

\section{Conclusion}
\label{sec:conclusion}

In this work, we experimentally demonstrated a Gaussian-modulated LLO CV-QKD system employing coherent dual-pilot-assisted frequency and phase recovery over 12~km of optical fiber. Operating at 156.25~MBaud with independent laser sources and RF-heterodyne detection, the proposed DSP architecture integrates CAZAC-based frame synchronization, global and subframe-wise CFO correction, coherent pilot combining, and matched filtering. By jointly exploiting two frequency-multiplexed pilot tones, the receiver compensates for laser-induced frequency and phase fluctuations, enabling reliable recovery of Gaussian-modulated quantum symbols.

The experimental analysis comprised 120 independently processed frames, each containing approximately $10^6$ coherent states. The system achieved an average channel transmittance of $0.5855\pm0.0081$ and input-referred excess noise of $15.56\pm7.61$~mSNU. Compared with single-pilot processing, the dual-pilot DSP chain reduced the average excess noise by 28.1\% and 17.4\% relative to pilots 1 and 2, respectively, consistently outperforming both configurations across all 120 frames. Under a trusted-receiver model, the average modeled secret-key rates reached $14.98\pm4.18$~Mbit/s in the asymptotic regime and $8.17\pm2.53$~Mbit/s when conservative finite-size parameter-estimation bounds were considered.

These results demonstrate the experimental benefits of jointly exploiting two pilot references for frequency and phase recovery in LLO CV-QKD systems. Although the observed improvements characterize the complete dual-pilot DSP chain, they highlight the potential of multi-pilot processing to mitigate excess noise in practical implementations. Future work will investigate operation at GBaud symbol rates, systematically compare coherent pilot combining with conventional phase averaging, and extend the security analysis toward composable finite-key guarantees. Further developments will also consider deployment over metropolitan fiber infrastructure, including potential integration into the Rede Rio Quântica Network.
\section*{Funding}
This work has been fully funded by the project \textit{Demonstração offline do sistema CV-QKD}, supported by QuIIN - Quantum Industrial Innovation EMBRAPII CIMATEC Competence Center in Quantum Technologies, with financial resources from the PPI IoT/Manufatura 4.0 of the MCTI grant number 053/2023, signed with EMBRAPII. 

\section*{Acknowledgments}
The authors acknowledge contributions from the NITeQ and QuIIN Testbed members, especially João M. Neto, Kamilla B. Rodrigues, João Vitor Fraga, Gabriel Almeida, Alexandro J. Snow, and B. Miranda.

\section*{Data availability}

The data and source code underlying the results presented in this work may be made available upon formal request, subject to review and approval by the authors.

\section*{Disclosures}
The authors declare that there are no conflicts of interest related to this article.

\bibliography{sample}

\end{document}